\documentclass[sigconf]{acmart}
\setcopyright{none}
\makeatletter
\newif\if@restonecol
\makeatother

\usepackage[linesnumbered,ruled,vlined]{algorithm2e}
\usepackage{algpseudocode}
\usepackage{amsmath}

\usepackage{enumitem}
\usepackage{multicol}
\usepackage{multirow}
\usepackage{float}
\usepackage{subfig}
\usepackage{balance}
\usepackage{bm}
\usepackage{array}
\usepackage[table]{xcolor}
\usepackage{tabularx}
\usepackage{booktabs}
\usepackage{makecell}
\usepackage{enumitem}
\usepackage[misc]{ifsym}

\newcommand{\siftTen}

\newcommand{\bfit}[1]{\textbf{\textit{#1}}}

\definecolor{deepPink}{RGB}{204,0,102}

\AtBeginDocument{%
  }

\setcopyright{acmlicensed}
\copyrightyear{2018}
\acmYear{2018}
\acmDOI{XXXXXXX.XXXXXXX}

\acmConference[Conference acronym 'XX]{Make sure to enter the correct
  conference title from your rights confirmation emai}{June 03--05,
  2018}{Woodstock, NY}

\acmISBN{978-1-4503-XXXX-X/18/06}

\begin{document}


\title{Vector Search for the Future: From Memory-Resident, Static Heterogeneous Storage, to Cloud-Native Architectures}



\author{Yitong Song}
\affiliation{
  \institution{Hong Kong Baptist University}
  \country{}
  }
\email{yitong\_song@hkbu.edu.hk}

\author{Xuanhe Zhou}
\affiliation{
  \institution{Shanghai Jiao Tong University}
  \country{}
  }
\email{zhouxuanhe@sjtu.edu.cn}

\author{Christian S. Jensen}
\affiliation{
  \institution{Aalborg University}
  \country{Denmark}
  }
\email{csj@cs.aau.dk}

\author{Jianliang Xu}
\affiliation{
  \institution{Hong Kong Baptist University}
  \country{}
  }
\email{xujl@comp.hkbu.edu.hk}







\renewcommand{\shortauthors}{Yitong Song et al.}


\begin{abstract}
Vector search (VS) has become a fundamental component in multimodal data management, enabling core functionalities such as image, video, and code retrieval. As vector data scales rapidly, VS faces growing challenges in balancing search, latency, scalability, and cost. The evolution of VS has been closely driven by changes in storage architecture. Early VS methods rely on all-in-memory designs for low latency, but scalability is constrained by memory capacity and cost. To address this, recent research has adopted heterogeneous architectures that offload space-intensive vectors and index structures to SSDs, while exploiting block locality and I/O-efficient strategies to maintain high search performance at billion scale. Looking ahead, the increasing demand for trillion-scale vector retrieval and cloud-native elasticity is driving a further shift toward memory–SSD–object storage architectures, which enable cost-efficient data tiering and seamless scalability. 

In this tutorial, we review the evolution of VS techniques from a storage-architecture perspective. We first review memory-resident methods, covering classical IVF-, hash-, quantization-, and graph-based designs. We then present a systematic overview of heterogeneous storage VS techniques, including their index designs, block-level layouts, query strategies, and update mechanisms. Finally, we examine emerging cloud-native systems and highlight open research opportunities for future large-scale vector retrieval systems.
\end{abstract}



\maketitle

\section{Introduction}
\begin{figure}[!t]
    \centering
    \includegraphics[width=\linewidth]{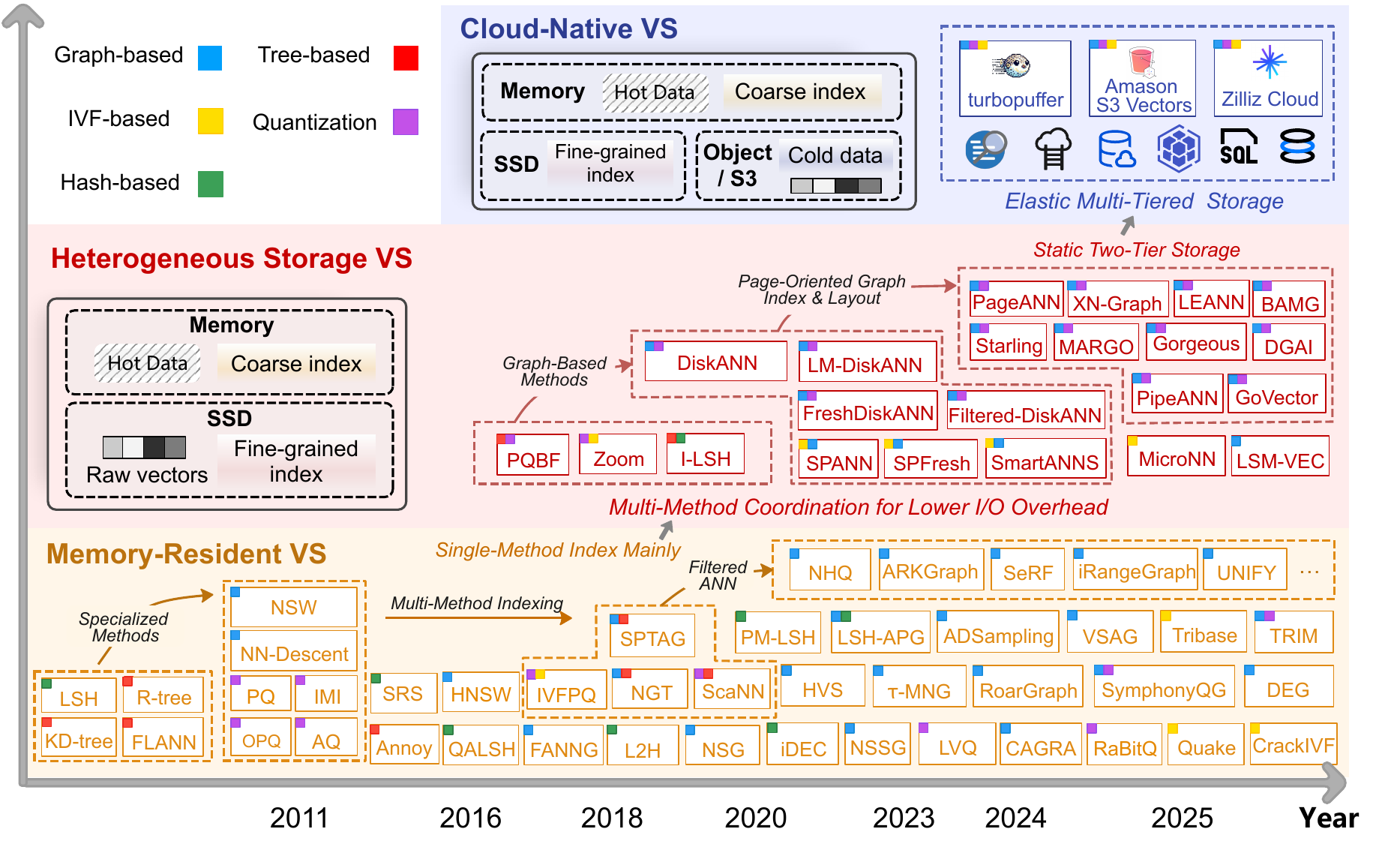}
    \caption{The Evaluation of Vector Search Techniques.}
    \label{fig:trends}
\end{figure}

Vector search (VS) aims to retrieve semantically similar items in a high-dimensional embedding space, and has become a fundamental building block for large-scale multimodal data search, supporting applications such as image-to-image retrieval in Taobao~\cite{ADBV}, similar-video retrieval in TikTok~\cite{BlendHouse}, and code assistants like Trae~\cite{gao2025trae}.
As application demands grow from million- to billion- and now trillion-scale, the storage of vector data and index structures has evolved from purely in-memory designs to heterogeneous memory–SSD deployments and further toward cloud-native architectures. This shift in adopted storage architecture has driven a significant evolution in VS techniques (see Figure~\ref{fig:trends}).

\noindent\bfit{(1) Memory-Resident VS.} Early VS methods store both raw vectors and index structures entirely in memory, enabling extremely high performance by optimizing computations while largely ignoring data access costs. Under this setting, a variety of techniques have been developed, including tree-based~\cite{ScalableNN, annoy}, IVF-based~\cite{meta-faiss, Tribase, CrackIVF, Quake}, hash-based~\cite{QALSH, srs, idec, TODS, learnToHash}, quantization-based~\cite{pq, OPQ, RaBitQ, CacheLocality, vectorCompression, VQ, IMI}, and graph-based approaches~\cite{HNSW, NSG, tau-MNG, HVS, LSH-APG, UNIFY}.
However, as vector volumes grow and memory remains costly, fully storing both vectors and memory-intensive indexes becomes impractical.

\noindent\bfit{(2) Static Heterogeneous-Storage VS.} To address the scalability and cost bottlenecks of memory-resident VS, a range of heterogeneous-storage VS methods have been developed. Initiated by PQBF in 2017~\cite{pqbf} and expanded rapidly since 2024 with methods such as Starling~\cite{starling}, PipeANN~\cite{PipeANN}, and PageANN~\cite{PageANN}, most approaches adopt a static two-tier memory–SSD architecture. In this design, space-intensive raw vectors and fine-grained index structures are placed on SSDs, while compressed vectors or coarse navigational structures are kept in memory. Building on this architecture, recent methods further exploit SSD-friendly layouts, block-level locality, and I/O-efficient search strategies to reduce data-access overhead and deliver low-latency performance. 

\noindent\bfit{(3) Elastic Multi-Tiered VS}. As vector data approach trillion scale, the static memory–SSD architecture faces cost–performance trade-offs and lacks elasticity. This has motivated a shift toward elastic multi-tiered designs (particularly in cloud-native scenarios~\cite{CloudNativeVS,ZillizCloud}) that integrate memory, SSD, and object/S3 storage under a unified, low-overhead architecture. Modern systems such as Zilliz Cloud~\cite{ZillizCloud}, Amazon S3 Vectors~\cite{AmazonS3}, and Turbopuffer~\cite{turbopuffer} adopt this model by dynamically placing vectors and index components across tiers according to access patterns and cost-performance considerations. Object storage provides virtually unlimited capacity and scales independently of compute resources, enabling seamless expansion, while frequently accessed data is promoted to warmer tiers (e.g., SSD or memory) on demand. This elastic, workload-aware migration strategy makes multi-tiered VS well-suited for trillion-scale vector search scenarios.



\begin{figure}[!t]
    \centering
    \includegraphics[width=0.95\linewidth]{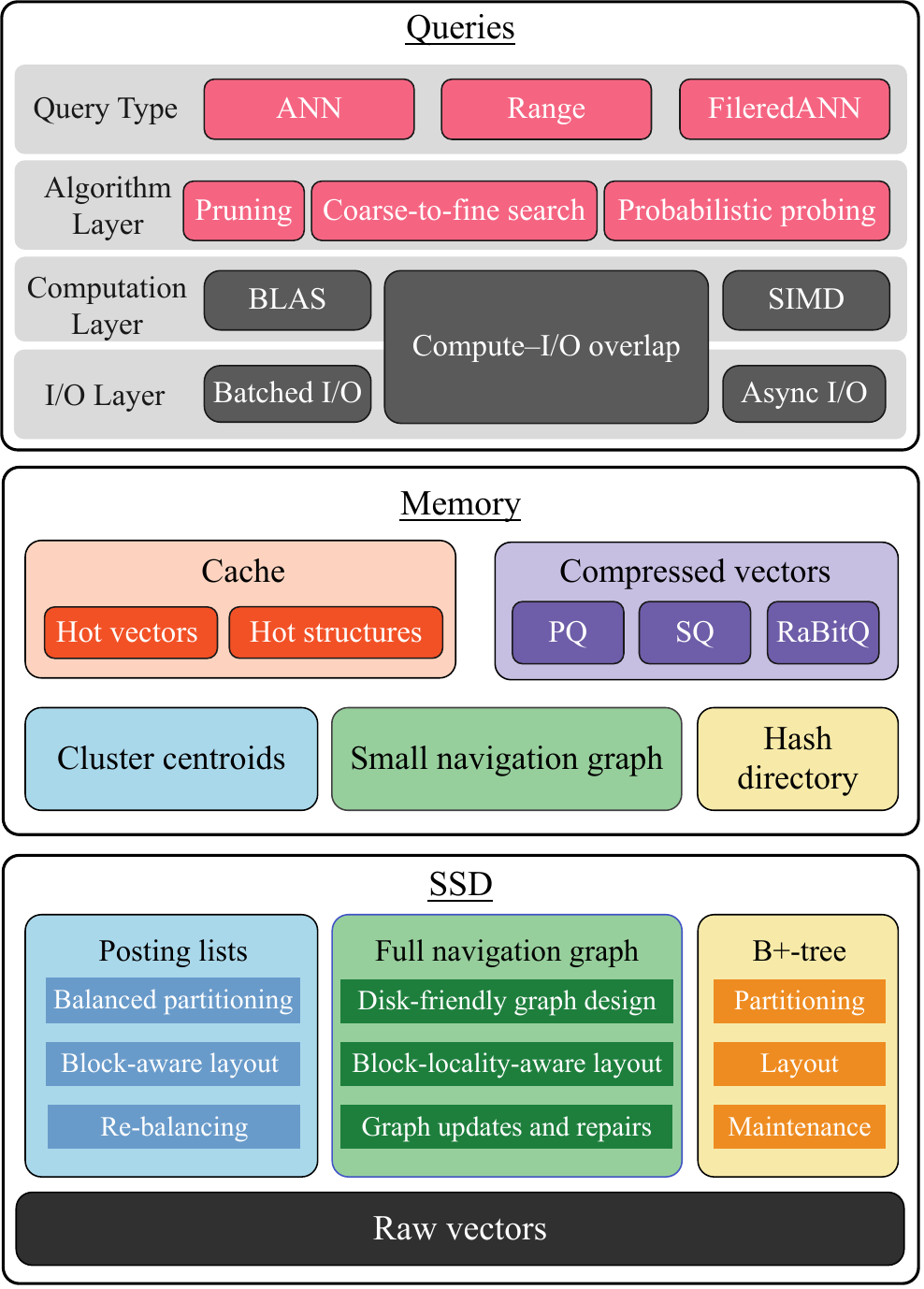}
    \caption{An Overview of Hetero.-Storage VS Techniques.}
   \vspace{-0.1in}
    \label{fig:tech}
\end{figure}

Motivated by these developments, this tutorial presents a comprehensive overview of VS techniques developed under the three storage architectures. We begin with a brief review of memory-resident methods, followed by an in-depth discussion of the rapidly expanding set of heterogeneous-storage VS techniques. Finally, we highlight emerging cloud-native practices and outline future research opportunities enabled by elastic multi-tiered architectures.

\noindent
\textbf{Tutorial Overview.} This 3-hour tutorial is organized into three parts based on the storage  architectures.

\noindent
\textbf{Part 1 (0.5 hours): Memory-Resident VS.} 
We briefly review classical in-memory VS methods and introduce how representative IVF-based, hash-based, quantization-based, and graph-based techniques work under a fully memory-resident setting, providing the technical foundations for subsequent sections.

\noindent
\textbf{Part 2 (2 hours): Heterogeneous-Storage VS.}
This is the main part of the tutorial. We present a unified technique architecture (Figure~\ref{fig:tech}), which captures common design patterns in modern memory-SSD methods. To provide a clear and structured exposition, we organize existing studies into IVF-, graph-, and tree-based categories, and discuss the techniques within each category.

\noindent
(1) \textit{IVF-based methods (40 min).} These methods partition vectors into clusters and restrict query processing to
probing only a small number of relevant clusters. We discuss four key components: (a) clustering and partitioning, (b) block-aware posting-list layouts, (c) probing-based query execution and optimization, and (d) update and maintenance mechanisms.

\noindent
(2) \textit{Graph-based methods (60 min).} These approaches organize vectors into proximity graphs and answer queries by traversing these graphs. We present these techniques along four aspects: (a) disk-friendly graph design, (b) block-locality-aware layouts, (c) coarse-to-fine search and optimizations, and (d) graph updates and repairs.

\noindent
(3) \textit{Tree-based methods (20 min).} These approaches organize vectors into disk blocks guided by directory structures such as B+-trees, often combined with hash functions or learned models to map high-dimensional vectors into one-dimensional keys. We highlight three aspects: (a) block-oriented partitioning and layouts, (b) block filtering and query execution, and (c) update mechanisms.

\begin{sloppypar}
\noindent
\textbf{Part 3 (0.5 hours): Elastic Multi-Tiered VS.}
We highlight two emerging VS systems, Ziliz Cloud~\cite{ZillizCloud} and Turbopuffer~\cite{turbopuffer}, that adopt multi-tiered storage, and discuss future research directions.
\end{sloppypar}

\noindent
\textbf{Importance to Data Management Community.} As databases increasingly incorporate vector data to support embedding-based search, VS is rapidly becoming a first-class workload in modern DBMSs. However, the community still lacks a comprehensive understanding of how vector data should be stored, indexed, and updated efficiently at scale. This tutorial provides practical guidance and fundamental design principles for VS development by clarifying: (1) how different storage tiers reshape the design of VS methods; (2) how memory-SSD VS methods are designed to address the dominant I/O bottlenecks; and (3) how cloud-native VS systems both benefit from and are challenged by tiered storage architectures.

\noindent
\textbf{Target Audience and Learning Outcomes.} This tutorial is intended for SIGMOD attendees from both academia and industry who are interested in vector retrieval, vector databases, or multimodal retrieval systems. We start with in-memory VS techniques, using intuitive explanations and motivating examples to build foundational concepts from scratch, and then progressively extend to memory-SSD and multi-tiered storage architectures. The material is self-contained, and no prior background is required. Attendees will learn how storage architectures shape VS methods and gain practical principles for building scalable and efficient VS systems.

\noindent
\textbf{Difference with Existing Tutorials.} Existing tutorials mainly focus on in-memory VS algorithms~\cite{tutorial2} or practical vector DBMSs\cite{tutorial1}. 
However, since these tutorials, a wave of research~\cite{starling, margo, Gorgeous, PipeANN, PageANN, XN-Graph, LEANN, BAMG, DGAI, turbopuffer, ZillizCloud} has driven VS beyond the in-memory paradigm, introducing heterogeneous-storage designs and elastic multi-tiered systems with new bottlenecks and design principles that were not the focus of earlier tutorials.
This tutorial responds to these shifts by adopting a storage-architecture-centric perspective, systematically tracing the evolution of VS techniques from memory-resident designs to heterogeneous memory–SSD solutions and further to cloud-native multi-tiered architectures. We provide an in-depth discussion on how memory–SSD methods address dominant I/O bottlenecks and how elastic multi-tiered VS is architected.














\begin{figure*}
    \centering
    \includegraphics[width=0.8\linewidth]{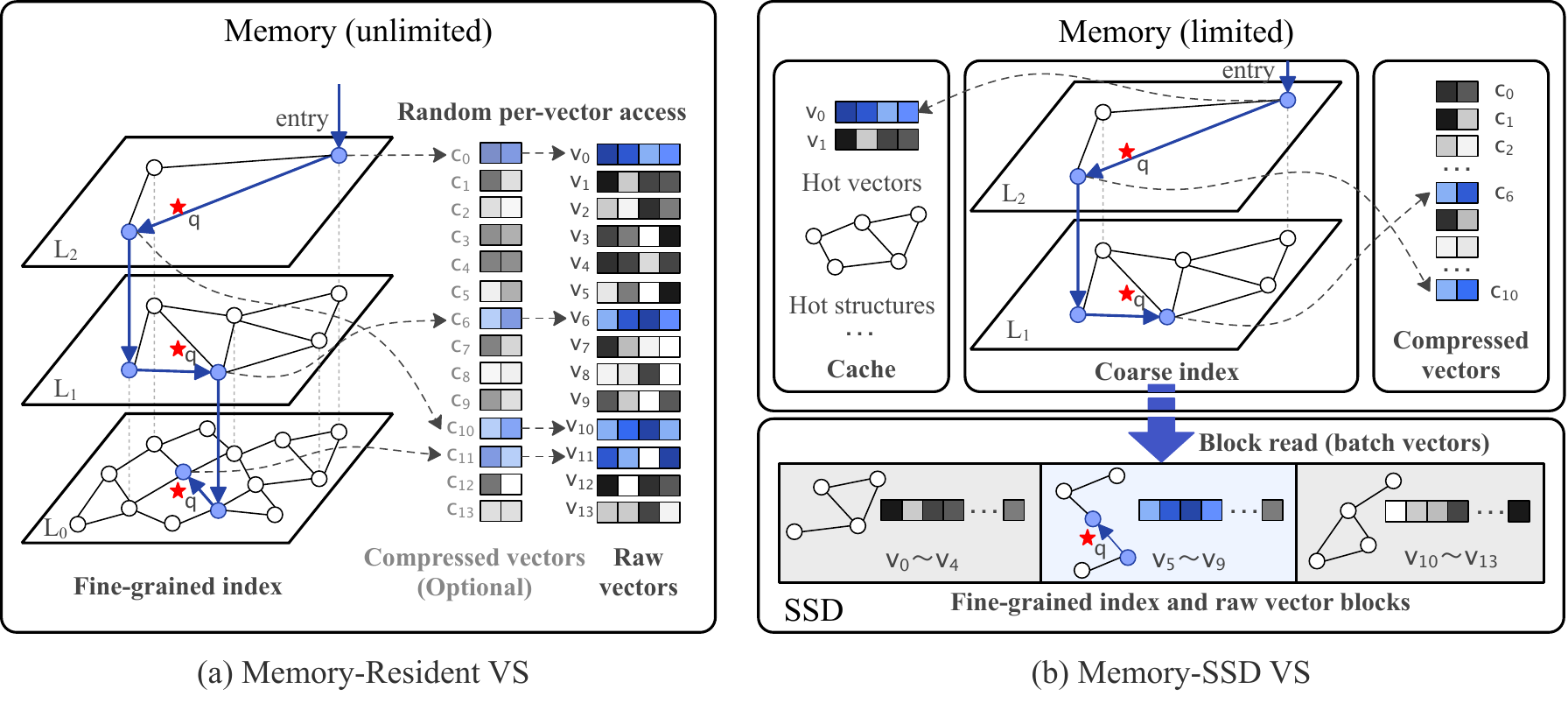}
    \vspace{-0.3cm}
    \caption{Memory-Resident VS vs. Memory-SSD VS (illustrated with a graph-based index).}
    \vspace{-0.3cm}
    \label{fig:diff}
\end{figure*}

\section{Tutorial Outline}
\label{sec: outline}
This section outlines the tutorial contents and structure, which follows the evolution of VS across three storage architectures: memory-resident, heterogeneous storage, and elastic multi-tiered designs.

\subsection{Memory-Resident VS} 
Memory-resident VS methods store both raw vectors
and index structures entirely in memory, thereby enabling low-latency data access and rapid query execution. We review four representative classes of in-memory methods: IVF-based, hash-based, quantization-based, and graph-based. 

\noindent
\textbf{IVF-based methods.} Inverted File (IVF) techniques organize vectors into clusters using algorithms such as $k$-means~\cite{K-means} and assign each vector to its nearest centroid. Each cluster is associated with an inverted list 
that stores the raw vectors or their compressed representations. During querying, the query vector first identifies $nprobe$ most relevant clusters by comparing it with the centroids, and only vectors within those clusters are further evaluated. IVF is simple to implement and easy to support updates. However, {\it it performs only coarse-grained pruning, since the probed clusters may still contain many vectors to scan.}

\noindent
\textbf{Hash-based methods.} Hash-based methods, particularly locality-sensitive hashing (LSH), map vectors into buckets using hash functions designed such that similar vectors are likely to collide in the same bucket. At query time, the query vector is hashed in the same way,  so that search is restricted to buckets with similar or identical hash codes. This approach provides theoretical guarantees for VS: the returned results are provably close to the ground truth within a bounded error, and the query time can be asymptotically lower than that of a linear scan, making these methods theoretically attractive. However, {\it in practice, achieving high accuracy with hash-based methods is challenging, as it often requires long hash codes and multiple hash tables, which lead to sparse buckets, excessive memory usage, and degraded performance.}

\noindent
\textbf{Quantization-based methods.} These methods compress raw vectors into compact codes, enabling distances between a query vector and data vectors to be estimated from these codes at much lower computational cost. Among them, Product Quantization (PQ)~\cite{pq} is one of the most widely used techniques: it splits each vector into multiple low-dimensional subspaces and, in each subspace, learns a codebook by clustering sub-vectors; each sub-vector is then encoded by the index of its closest centroid. During querying, approximate distances are obtained using only the compact codes, thereby avoiding expensive operations on raw vectors. This approach offers strong memory efficiency and fast distance evaluation. However, {\it the compression inevitably introduces approximation error, which can degrade query accuracy.}

\noindent
\textbf{Graph-based methods.} These methods connect each vector to its approximate nearest neighbors in diverse directions, forming proximity graphs that support efficient greedy navigation. A query starts from one or several entry points and iteratively explores neighbors that bring it closer to the target vector, stopping when no candidate vector offers further improvement. Representative structures such as HNSW~\cite{HNSW} further enhance query efficiency by organizing vectors into hierarchical graphs: the upper layers contain progressively sparser graphs that guide the search quickly toward a region near the true nearest neighbors, while the lower layers provide fine-grained retrieval for refinement. This hierarchical navigation substantially shortens the search path and reduces query latency. Graph-based methods generally offer the best accuracy–efficiency trade-off among in-memory VS techniques and dominate many high-performance applications~\cite{kANNSurvey, graphSurvey}. {\it However, these graphs are memory-intensive, and their search process involves substantial random data access, making it challenging to extend them to disk-resident settings.}

\begin{table*}
\centering
\caption{Comparison of Representative Memory-SSD VS Methods.}
\label{tab:summary}
{
\footnotesize
\renewcommand{\arraystretch}{1.2}
\renewcommand{\tabularxcolumn}[1]{>{\centering\arraybackslash}m{#1}}
\setlength{\tabcolsep}{4pt}
\begin{tabularx}{\textwidth}{%
    |>{\centering\arraybackslash}m{0.9cm}|
    c|
    >{\centering\arraybackslash}m{2.6cm}|
    >{\centering\arraybackslash}m{2.6cm}|
    >{\centering\arraybackslash}m{2.8cm}|
    >{\centering\arraybackslash}m{0.55cm}|
    >{\centering\arraybackslash}m{0.4cm}|
    >{\centering\arraybackslash}m{0.7cm}|
    >{\centering\arraybackslash}m{0.5cm}|
    >{\centering\arraybackslash}m{2.15cm}|
}
\cline{1-10}
\cline{1-10}
\multirow{2}{*}{\textbf{Type}} &
\multirow{2}{*}{\textbf{Method}} &
\multirow{2}{*}{\textbf{Key Tech.}} &
\multicolumn{2}{c|}{\textbf{Storage}} &
\multicolumn{3}{c|}{\textbf{Query Type}} &
\multirow{2}{*}{\makecell[c]{\textbf{Up.-}\\\textbf{Opt.}}} & \multirow{2}{*}{\textbf{Applications}} \\
\cline{4-8}
\cline{4-8}
& & &\centering DRAM &\centering SSD &\centering ANN &\centering RS &\centering FANN & & \\
\cline{1-10}
\cline{1-10}

\multirow{4}{*}{\vspace{-.4cm} \makecell[c]{IVF-\\based}}
& SPANN~\cite{spann} & Hierarchical Balanced Clustering & Centroids & Posting Lists + Vectors & $\checkmark$ & $\times$ & $\times$ & $\times$ & \multirow{4}{*}{\vspace{-0.5cm}\makecell[c]{Dynamic \\workloads}} \\
\cline{2-9}
& SPFresh~\cite{spfresh} & Lightweight Rebalancing & Centroids & Posting Lists + Vectors & $\checkmark$ & $\times$ & $\times$ & $\checkmark$ & \\
\cline{2-9}
& MicroNN~\cite{MicroNN} & On-device Lightweight Cluster Reorganization & Centroids & Posting Lists + Vectors & $\checkmark$ & $\times$ & $\times$ & $\checkmark$ & \\
\cline{2-9}
& Zoom~\cite{zoom} & In-Memory Approximate Search \& SSD Exact Reranking & Compact Vectors + Centroids + Posting Lists & Vectors & $\checkmark$ & $\times$ & $\times$ & $\times$ & \\
\cline{1-10}

\multirow{6}{*}{\vspace{-1cm}\makecell[c]{Graph-\\based}}
& DiskANN~\cite{diskann} & Vamana graph & Compact Vector & Vamana + Vectors  & $\checkmark$ & $\times$ & $\times$ & $\times$ & \multirow{6}{*}{\vspace{-1cm}\makecell[c]{Latency-critical, \\ high-accuracy \\ search}}\\
\cline{2-9}
& \makecell[c]{Filtered-\\DiskANN~\cite{Filtered-diskann}}
 & Vamana with Filtering Support & Compact Vector & Vamana + Vectors & $\checkmark$ & $\times$ & $\checkmark$ & $\times$ & \\
\cline{2-9}
& 
\makecell[c]{Fresh-\\DiskANN~\cite{FreshDiskANN}} & Dynamic Update Strategies & Compact Vector & Vamana + Vectors & $\checkmark$ & $\times$ & $\times$ & $\checkmark$ & \\
\cline{2-9}
& Starling~\cite{starling} & Optimized Layouts & Compact Vector + Entry Graph & Vamana + Vectors & $\checkmark$ & $\checkmark$ & $\times$ & $\times$ & \\
\cline{2-9}


\cline{2-9}
& LSM-VEC~\cite{LSM-VEC} & HNSW + LSM & Upper Layer Graphs & Bottom Layer Graph & $\checkmark$ & $\times$ & $\times$ & $\checkmark$ & \\
\cline{2-9}
& PageANN~\cite{PageANN} & Page-Aware Layouts & Compact Vector + Entry Graph & Graph + Vectors & $\checkmark$ & $\times$ & $\times$ & $\times$ & \\ 
\cline{2-9}

\cline{1-10}
\multirow{3}{*}{
\vspace{-0.5cm} 
\makecell[c]{Tree-\\based}}
& \makecell[c]{I-LSH\cite{I-LSH} / \\ EL-LSH~\cite{ei-lsh}} & Query-Aware LSH & Hash Tables & Multiple $\text{B}^+$-Tree & $\checkmark$ & $\times$ & $\times$ & $\times$ & \multirow{3}{*}{
\vspace{-0.5cm} 
\makecell[c]{Low-to-medium\\dimensions}} \\
\cline{2-9}
& NeOPFA~\cite{icde_learned_functions} 
& Learned Sorted-Lists with Sequential Page Access 
& Hash Tables   
& Sorted Lists (indexed by $\text{B}^+$-tree) + Vectors 
& $\checkmark$ 
& $\times$ 
& $\times$ 
& $\times$ 
& \\
\cline{2-9}
&PQBF~\cite{pqbf} 
&PQ-based Filtering with Tree Pruning
&PQ Metadata + Directory 
&Compact Vectors+ PQB$^+$-Tree 
&$\checkmark$ 
&$\times$ 
&$\times$ 
&$\times$ 
& \\
\cline{1-10}
\end{tabularx}
\begin{flushleft}
* RS, FANN, and Up.-Opt. denote range search, filtered ANN search (ANN with attribute predicates), and whether the method offers optimized update support, respectively.\\
\end{flushleft}
}
\end{table*}

\subsection{Heterogeneous-Storage VS}
To support billion-scale deployments while preserving low-latency search, recent methods adopt a heterogeneous memory–SSD storage architecture, where memory stores lightweight structures for routing and coarse pruning, and SSDs hold space-intensive raw vectors and fine-grained indexes. Unlike memory-resident VS, which can rely on a single index paradigm and focus on reducing computational cost, heterogeneous-storage methods often combine multiple complementary indexes (e.g., quantization + graph, hashing + tree) and introduce new challenges to minimizing random I/O.

In this tutorial, we categorize existing heterogeneous memory–SSD solutions by their core fine-filtering structures, i.e., IVF-based, graph-based, and tree-based methods, and present the corresponding design principles and optimization strategies. Table~\ref{tab:summary} summarizes representative methods in each category along with the query types and application scenarios they target.

\subsubsection{IVF-Based Methods.}  IVF-based VS methods on heterogeneous memory–SSD storage extend the classical in-memory IVF paradigm by keeping cluster centroids and lightweight routing structures in memory, while storing raw (or compressed) vectors and full posting lists on SSD. The key challenge lies in designing balanced clustering and SSD-aware layouts so that probing can be served by a small number of locality-preserving block reads rather than costly scattered random I/O. We introduce the related techniques through four key components: (i) clustering and partitioning, which assign vectors to balanced clusters to maintain bounded list sizes and enhance co-access locality; (ii) block-aware posting-list layout, which aligns and groups list segments to improve page efficiency and reduce cross-block fetches; (iii) probing-based query execution and optimizations, which control the number of probed clusters and employ techniques to further lower latency; and (iv) update and maintenance mechanisms, which preserve cluster balance and layout locality under frequent updates.

\noindent
\textbf{Clustering and partitioning.} Most methods employ balanced $k$-means~\cite{BalancedKMeans, BalancedCluster} to control list sizes or mini-batch $k$-means~\cite{MicroNN, meta-faiss} to operate under limited memory. Balanced clustering is particularly important in SSD-resident settings, as it helps posting lists align with SSD blocks and avoids oversized clusters that would cause excessive page fetches during probing. SPANN~\cite{spann} further adopts hierarchical balanced clustering to refine routing in a coarse-to-fine manner, thereby reducing the probed cluster count and improving data locality. 

\noindent
\textbf{Block-aware posting-list layout.} A key challenge in SSD-resident IVF is organizing posting lists to minimize I/O. Common strategies include block-aligned list segments~\cite{meta-faiss}, locality-enhanced contiguous layouts that group similar clusters~\cite{spann, MicroNN}, and segmentation of posting lists into hot and cold regions~\cite{CloudNativeVS}. For FANN workloads, Hybrid-IVFFlat~\cite{hy_ivf_flat} colocates vector identifiers and filterable attributes within each posting-list entry, enabling filtering and candidate pruning in a single sequential pass without additional random reads. Moreover, its posting lists are further divided into independently loadable segments, so only segments relevant to the filter predicates need to be fetched. 

\noindent
\textbf{Probing-based query execution and optimization.} During querying, the system fetches the top-$nprobe$ clusters whose centroids are closest to the query. Choosing $nprobe$ is critical: too small a value harms recall, while too large a value increases I/O cost. To reduce latency, recent methods employ {\it adaptive list selection}, {\it batched I/O}, {\it prefetching}, and {\it compute–I/O overlap} to make cluster access more efficient. Zoom~\cite{zoom} employs a preview–reranking strategy in which compact vector previews filter candidates before fetching SSD-resident vectors, reducing both I/O volume and tail latency. SPANN~\cite{spann} uses hierarchical clustering to restrict probing to localized partitions, lowering the number of accessed lists per query.

\noindent
\textbf{Update and maintenance mechanisms.} IVF-based methods naturally support for dynamic updates. Insertions can be handled by simply appending vectors to the corresponding posting lists, and deletions require only lightweight list modifications. However, frequent updates may skew cluster sizes and reduce query efficiency. To address this, MicroNN~\cite{MicroNN} introduces a \emph{delta-store} that temporarily buffers new or updated vectors and periodically merges it into the main posting lists in the background, avoiding immediate disruption to SSD-resident data. SPFresh~\cite{spfresh} focuses on maintaining cluster balance by applying \emph{lightweight rebalancing} that selectively migrates only a few vectors between clusters, preserving partition quality without requiring costly full re-clustering.

\subsubsection{Graph-Based Methods.} Graph-based VS methods extend in-memory graph indexes (e.g., HNSW~\cite{HNSW}, Vamana~\cite{diskann}) to heterogeneous memory–SSD architectures by retaining lightweight navigational structures and compact vectors in memory, while storing full adjacency lists and raw vectors on SSD. This shift introduces a fundamental new challenge: each neighbor expansion may trigger a page fetch, and thus the graph must be redesigned so that traversal can proceed in block-sized batches with strong locality. In contrast to in-memory graphs, where data access is almost free, SSD-resident graphs must carefully reshape both graph structures and physical layouts to minimize I/O without sacrificing accuracy. We categorize existing techniques into four core components: (i) disk-friendly graph design, which enforces degree bounds and directional diversity so that adjacency lists can align with SSD blocks; (ii) block-locality-aware layouts, which organize neighbors and raw vectors into locality-preserving SSD blocks; (iii) coarse-to-fine search and optimizations, which employ batched I/O, prefetching, and compute–I/O overlap to reduce latency; and (iv) graph update and repair strategies, which maintain graph quality and block locality under dynamic workloads.

\noindent
\textbf{Disk-friendly graph design.} Unlike memory-resident graphs, SSD-resident graphs must maintain small, locality-preserving, and block-aligned adjacency lists to reduce random I/O. Representative designs such as Vamana~\cite{diskann}, XN-Graph~\cite{XN-Graph}, BAMG~\cite{BAMG}, and PageANN~\cite{PageANN} enforce degree control and edge diversification to keep graphs both navigable and SSD-friendly. Vamana applies occlusion-based pruning with a progress parameter to retain long-range edges and ensure fixed degrees suitable for page alignment. XN-Graph expands neighborhood coverage using eXtended Neighborhood Descent before applying occlusion-based pruning, enabling better long-range connectivity and improved navigability on SSD. BAMG distinguishes intra-/cross-block links to favor edges that provide progress without extra page fetches. PageANN further aggregates vectors into page-level nodes and removes redundant intra-page links to match physical page layouts. 

\noindent
\textbf{Block-locality-aware layouts.} A key design challenge in SSD-resident graphs is how to place adjacency lists, raw vectors, and metadata onto SSD blocks to minimize I/O. Existing methods fall into two broad categories: coupled layouts and decoupled layouts. Coupled layouts~\cite{diskann, starling, margo, PageANN} colocate a vector and its neighbors within the same page, so that one read supports both vector evaluation and neighbor expansion. Decoupled layouts~\cite{trim, BAMG, DGAI} store graph structures and raw vectors separately, so that each can be optimized for its own access pattern, graph index components are frequently accessed, while full vectors are only needed during final reranking.
However, a naïve neighbor list layout based on insertion or ID order does not guarantee that graph-near neighbors land in the same page, leading to excessive page hopping during traversal. To address this, recent work formulates block layout as a locality optimization problem that tightly groups mutually reachable neighbors within the same pages. DiskANN++~\cite{DiskANNplus} reorganizes nodes using star/bin packing to keep graph-close vertices within page-aligned units; Starling~\cite{starling} and MARGO~\cite{margo} maximize intra-page neighbor overlap so that each fetch yields more usable neighbors; and PageANN~\cite{PageANN} further enhances this idea by clustering similar vectors into page-level “super-nodes,” aligning logical graph navigation exactly with physical SSD pages and removing redundant intra-page edges.

\noindent
\textbf{Coarse-to-fine search and optimizations.} SSD-resident graph search still follows a best-first paradigm, but unlike in-memory traversal where neighbors are instantly accessible, each node expansion may incur expensive SSD reads. To mitigate this, recent systems adopt a coarse-to-fine search strategy: inexpensive in-memory structures (e.g., entry graphs, compact vectors) are first used to guide navigation toward promising regions, and fine-grained SSD accesses occur only when necessary. Starling~\cite{starling} demonstrates this by using a small in-memory entry graph for fast initial routing before touching SSD-resident neighbors, while PageANN~\cite{PageANN} pushes pruning to the page-level traversal, loading only those pages whose aggregated vectors are likely to contribute to the final results, thus improving the utility of each read. To further mask I/O latency, the search algorithm is also paired with asynchronous I/O, prefetching, and compute–I/O overlap, as well as lightweight pruning (e.g., beam-size control~\cite{starling}, lower-bound filtering~\cite{trim}) reduces the number of nodes expanded per page. 

\noindent
\textbf{Graph update and repair strategies.} Maintaining high performance under dynamic workloads is particularly challenging for SSD-resident graphs, since insertions and deletions can quickly degrade both navigability and block locality. FreshDiskANN~\cite{FreshDiskANN} inserts new vectors into a small in-memory dynamic graph and a dedicated fresh SSD region, while deletions are recorded as tombstones. A background streaming merge periodically integrates these updates into the main graph using a delete–insert–patch pipeline.
LSM-VEC~\cite{LSM-VEC} introduces a LSM-like index structure, where recent updates are handled in a fast in-memory layer and gradually merged into on-disk graph structures, thereby minimizing block rewrites and preserving disk-layout locality. 
DGAI~\cite{DGAI} uses a decoupled on-disk layout, so updating adjacency lists or vector data does not require rewriting entire pages. 

\subsubsection{Tree-Based Methods.} These methods organize vectors into SSD blocks through directory structures such as B\textsuperscript{+}-trees, learned indexes, or hash-based partitioning trees. A high-dimensional vector is first mapped into a one-dimensional key, typically via LSH, learned models, or quantization-based encodings, and these keys are then stored in sorted disk blocks. Query processing then involves probing only a small number of relevant blocks, avoiding large posting-list scans. Existing methods can be understood through three major components: (i) block-oriented partitioning and layouts, which determine how vectors are transformed into keys and organized into SSD blocks; (ii) block filtering and query execution, which govern how the system selects and probes relevant blocks during search; and (iii) update and maintenance mechanisms, which address how the index remains efficient under insertions, deletions, and data distribution changes. 

\noindent
\textbf{Block-oriented partitioning and layouts.} Building on the general idea of mapping vectors into a one-dimensional keys, different methods adopt distinct strategies to construct keys and organize SSD blocks for efficient access. I-LSH~\cite{I-LSH} derives keys from LSH bucket identifiers and arranges vectors into B\textsuperscript{+}-tree–indexed buckets, each stored as a contiguous block range. PQBF~\cite{pqbf} uses PQ codes to compute tight lower bounds on distances and organizes SSD blocks such that PQ-coded vectors belonging to similar coarse partitions are colocated. During querying, blocks are ranked by PQ estimated distances to the query and probed in distance order, enabling early pruning without scanning all blocks.
The learned-function method~\cite{icde_learned_functions} trains regression models to map vectors to monotonic keys that better preserve global ordering and then materialize SSD blocks as sorted runs with high locality.

\noindent
\textbf{Block filtering and query execution.} Query processing begins by mapping the query vector into the same key space and identifying promising block ranges to probe. I-LSH~\cite{I-LSH} performs probabilistic range probing around the query’s hash-derived key, while EI-LSH~\cite{ei-lsh} introduces incremental probing with early termination, gradually expanding the search radius until sufficient candidates are found. PQBF~\cite{pqbf} improves block filtering by ordering blocks based on PQ-approximated distances, enabling distance-ordered probing rather than relying purely on key ranges. Moreover, recent methods also combine batched I/O, prefetching, and compute–I/O overlap to reduce latency and amortize SSD access costs. 

\noindent
\textbf{Update and maintenance mechanisms.} Tree-based VS methods naturally support updates through the underlying B\textsuperscript{+}-tree or learned-directory structure: insertions place new vectors into the appropriate key range, and deletions remove or mark entries without restructuring the entire index. However, frequent updates may cause block imbalance, key skew, or degraded locality. 
EI-LSH~\cite{ei-lsh} addresses this by maintaining buffer zones and performing periodic block consolidation to limit fragmentation. The learned-function approach~\cite{icde_learned_functions} relies on incremental model refinement to maintain mapping accuracy when the data distribution drifts. When PQ-based partitioning is used~\cite{pqbf}, lightweight codebook retraining or background refinement helps preserve index quality. 

\subsection{Elastic Multi-Tiered VS}
With the rapid growth of vector volumes, enterprise VS deployments are increasingly embracing multi-tiered storage architectures that combine high-performance memory, large-capacity SSDs, and cost-efficient object stores. This tiered design allows systems to scale from billions to trillions of vectors while balancing latency, capacity, and cost efficiency. In this part, we present two representative industrial practices, i.e., Zilliz Cloud~\cite{ZillizCloud} and TurboPuffer~\cite{turbopuffer}, and discuss future research directions.

\noindent
\textbf{Zilliz Cloud.} Zilliz Cloud (built on Milvus 2.x) follows a memory–SSD–object architecture tiered by access frequency. Memory holds routing metadata, index headers, and recently accessed vectors to support ultra-low-latency lookups. SSDs persist frequently queried index files (e.g., HNSW/IVF/PQ) and recent update logs. Object storage serves as the elastic cold tier, storing raw vector segments and historical versions. Querying proceeds in a tier-aware pipeline: memory metadata locates target index shards, SSD-resident indexes retrieve candidates, and object-stored vectors are fetched on demand and gradually promoted to fast tiers. Updates follow an append-friendly model using delta logs and tombstones, while background compaction periodically merges them to preserve index quality without full rebuilds.

\noindent
\textbf{TurboPuffer.} TurboPuffer adopts a similar multi-tier optimized for ultra-large vector repositories. Hot data (recent inserts, active partitions, routing structures) remains in memory, whereas index structures and warm vectors are placed on SSDs for high-throughput querying. Cold partitions and historical vectors are stored in cloud object stores (e.g., S3/GCS), yet remain accessible via delayed pagination and speculative prefetching. Its update pipeline features streaming ingestion, batch commit, and lazy reorganization to mitigate fragmentation and maintain SSD locality. By integrating multi-tier caching, snapshot isolation, and cloud-native durability guarantees, TurboPuffer achieves scalable cost-efficient deployments with strong write performance.

\noindent
\textbf{Future Directions.} Although elastic multi-tiered VS methods have demonstrated strong potential in industrial systems, the field is still in its early stages of development. Many technical challenges remain open: (1) \emph{Tier-aware Index Co-Design}, where index construction and query strategies must jointly consider SSD- and object-storage-aware data layout and tier switching cost; (2) \emph{Adaptive and Predictive Caching}, which leverages access prediction to proactively migrate vectors and reduce latency; (3) \emph{Efficient Querying from Object Storage}, enabling progressive refinement and minimal on-demand I/O when accessing cold data; (4) \emph{Elasticity and Auto-Scaling}, supporting online repartitioning and performance-aware tier expansion under dynamic workloads; and (5) \emph{Cost Optimization}, jointly reducing compute overhead, SSD writes, and object-storage bandwidth while maximizing cost efficiency at cloud scale.




















\section{Biography}
\label{sec: bio}
\textbf{Yitong Song} is a  postdoctoral research fellow in the Department of Computer Science, Hong Kong Baptist University. Her research focuses on vector databases, with multiple publications at SIGMOD and VLDB. She will present heterogeneous-storage VS techniques.

\noindent
\textbf{Xuanhe Zhou} is an assistant professor in the Department of Computer Science, Shanghai Jiao Tong University. His research interests lie in data systems. He has received the SIGMOD 2025 Jim Gray Dissertation Honorable Mention and VLDB 2023 Best Industry Paper Runner-up. He has given tutorials at SIGMOD, VLDB, ICDE. He will present memory-resident VS techniques.

\noindent
\textbf{Christian S. Jensen} is a Professor of Computer Science at Aalborg University, Denmark. His research centers on temporal, spatial, and multimodal data management and analytics. An ACM Fellow and IEEE Fellow with an h-index of 105, he has published over 500 papers and received the 2022 ACM SIGMOD Contributions Award and 2019 IEEE TCDE Impact Award. Former Editor-in-Chief of ACM TODS and The VLDB Journal, he has delivered numerous influential tutorials at SIGMOD and VLDB. He will present the background and research opportunities.

\noindent
\textbf{Jianliang Xu} is a Chair Professor and Head of the Department of Computer Science, Hong Kong Baptist University. His research spans databases, blockchain, and privacy-aware data management. An IEEE Fellow with an h-index of 65, he has published over 300 papers and given tutorials at SIGMOD and ICDE. He will present elastic multi-tiered VS techniques and open problems.

\bibliographystyle{ACM-Reference-Format}
\balance
\bibliography{reference}

@String{Springer = "Springer-Verlag" }

@article{kANNSurvey,
  title={Approximate nearest neighbor search on high dimensional data—experiments, analyses, and improvement},
  author={Li, Wen and Zhang, Ying and Sun, Yifang and Wang, Wei and Li, Mingjie and Zhang, Wenjie and Lin, Xuemin},
  journal={IEEE TKDE},
  volume={32},
  number={8},
  pages={1475--1488},
  year={2019},
}

@misc{meta-faiss,
   author = {Meta AI},
   title = {FAISS},
   howpublished = {\url{https://ai.facebook.com/tools/faiss}},
   year = {2017},
 }

@article{ADBV,
  title={Analyticdb-v: A hybrid analytical engine towards query fusion for structured and unstructured data},
  author={Wei, Chuangxian and Wu, Bin and Wang, Sheng and Lou, Renjie and Zhan, Chaoqun and Li, Feifei and Cai, Yuanzhe},
  journal={PVLDB},
  volume={13},
  number={12},
  pages={3152--3165},
  year={2020},
}

@article{diskann,
  title={Diskann: Fast accurate billion-point nearest neighbor search on a single node},
  author={Jayaram Subramanya, Suhas and Devvrit, Fnu and Simhadri, Harsha Vardhan and Krishnawamy, Ravishankar and Kadekodi, Rohan},
  journal={NeurIPS},
  volume={32},
  year={2019}
}

@article{freshdiskann,
  title={FreshDiskANN: A Fast and Accurate Graph-Based ANN Index for Streaming Similarity Search},
  author={Singh, Aditi and Subramanya, Suhas Jayaram and Krishnaswamy, Ravishankar and Simhadri, Harsha Vardhan},
  journal={arXiv preprint arXiv:2105.09613},
  year={2021}
}

@inproceedings{Filtered-diskann,
  title={Filtered-diskann: Graph algorithms for approximate nearest neighbor search with filters},
  author={Gollapudi, Siddharth and Karia, Neel and Sivashankar, Varun and Krishnaswamy, Ravishankar and Begwani, Nikit and Raz, Swapnil and Lin, Yiyong and Zhang, Yin and Mahapatro, Neelam and Srinivasan, Premkumar and others},
  booktitle={WWW},
  pages={3406--3416},
  year={2023}
}

@article{starling,
  title={Starling: An I/O-Efficient Disk-Resident Graph Index Framework for High-Dimensional Vector Similarity Search on Data Segment},
  author={Wang, Mengzhao and Xu, Weizhi and Yi, Xiaomeng and Wu, Songlin and Peng, Zhangyang and Ke, Xiangyu and Gao, Yunjun and Xu, Xiaoliang and Guo, Rentong and Xie, Charles},
  journal={ACM SIGMOD},
  volume={2},
  number={1},
  pages={1--27},
  year={2024},
  publisher={ACM New York, NY, USA}
}

@inproceedings{spann,
  title={Spann: Highly-efficient billion-scale approximate nearest neighborhood search},
  author={Chen, Qi and Zhao, Bing and Wang, Haidong and Li, Mingqin and Liu, Chuanjie and Li, Zengzhong and Yang, Mao and Wang, Jingdong},
  journal={NeurIPS},
  volume={34},
  pages={5199--5212},
  year={2021}
}

@article{Tribase,
  title={Tribase: A Vector Data Query Engine for Reliable and Lossless Pruning Compression using Triangle Inequalities},
  author={Xu, Qian and Yang, Juan and Zhang, Feng and Pan, Junda and Chen, Kang and Shen, Youren and Zhou, Amelie Chi and Du, Xiaoyong},
  journal={ACM SIGMOD},
  volume={3},
  number={1},
  pages={1--28},
  year={2025},
  publisher={ACM New York, NY, USA}
}

@article{HNSW,
  title={Efficient and robust approximate nearest neighbor search using hierarchical navigable small world graphs},
  author={Malkov, Yu A and Yashunin, Dmitry A},
  journal={IEEE TPAMI},
  volume={42},
  number={4},
  pages={824--836},
  year={2018},
}

@inproceedings{NSG,
  title={Fast approximate nearest neighbor search with the navigating spreading-out graph},
  author={Fu, Cong and Xiang, Chao and Wang, Changxu and Cai, Deng},
  booktitle={PVLDB},
  volume = {12}, 
  number = {5},
  year={2019},
  pages={416--474},
  publisher={VLDB Endowment}
}

@article{tau-MNG,
  title={Efficient Approximate Nearest Neighbor Search in Multi-dimensional Databases},
  author={Peng, Yun and Choi, Byron and Chan, Tsz Nam and Yang, Jianye and Xu, Jianliang},
  journal={ACM SIGMOD},
  volume={1},
  number={1},
  pages={1--27},
  year={2023},
  publisher={ACM New York, NY, USA}
}

@article{HVS,
  title={HVS: hierarchical graph structure based on voronoi diagrams for solving approximate nearest neighbor search},
  author={Lu, Kejing and Kudo, Mineichi and Xiao, Chuan and Ishikawa, Yoshiharu},
  journal={PVLDB},
  volume={15},
  number={2},
  pages={246--258},
  year={2021},
  publisher={VLDB Endowment}
}

@article{LSH-APG,
  title={Towards Efficient Index Construction and Approximate Nearest Neighbor Search in High-Dimensional Spaces},
  author={Zhao, Xi and Tian, Yao and Huang, Kai and Zheng, Bolong and Zhou, Xiaofang},
  journal={PVLDB},
  volume={16},
  number={8},
  pages={1979--1991},
  year={2023},
  publisher={VLDB Endowment}
}

@article{QALSH,
  title={Query-aware locality-sensitive hashing for approximate nearest neighbor search},
  author={Huang, Qiang and Feng, Jianlin and Zhang, Yikai and Fang, Qiong and Ng, Wilfred},
  journal={PVLDB},
  volume={9},
  number={1},
  pages={1--12},
  year={2015},
}

@article{srs,
  title={SRS: solving c-approximate nearest neighbor queries in high dimensional euclidean space with a tiny index},
  author={Sun, Yifang and Wang, Wei and Qin, Jianbin and Zhang, Ying and Lin, Xuemin},
  journal={PVLDB},
  year={2014}
}

@article{idec,
  title={iDEC: indexable distance estimating codes for approximate nearest neighbor search},
  author={Gong, Long and Wang, Huayi and Ogihara, Mitsunori and Xu, Jun},
  journal={PVLDB},
  volume={13},
  number={9},
  year={2020}
}

@article{TODS,
  title={Efficient and accurate nearest neighbor and closest pair search in high-dimensional space},
  author={Tao, Yufei and Yi, Ke and Sheng, Cheng and Kalnis, Panos},
  journal={ACM TODS},
  volume={35},
  number={3},
  pages={1--46},
  year={2010},
  publisher={ACM New York, NY, USA}
}

@inproceedings{learnToHash,
  title={A general and efficient querying method for learning to hash},
  author={Li, Jinfeng and Yan, Xiao and Zhang, Jian and Xu, An and Cheng, James and Liu, Jie and Ng, Kelvin KW and Cheng, Ti-chung},
  booktitle={ACM SIGMOD},
  pages={1333--1347},
  year={2018}
}

@article{pq,
  title={Product quantization for nearest neighbor search},
  author={Jegou, Herve and Douze, Matthijs and Schmid, Cordelia},
  journal={IEEE TPAMI},
  volume={33},
  number={1},
  pages={117--128},
  year={2010},
}

@inproceedings{OPQ,
  title={Optimized product quantization for approximate nearest neighbor search},
  author={Ge, Tiezheng and He, Kaiming and Ke, Qifa and Sun, Jian},
  booktitle={IEEE TPAMI},
  pages={2946--2953},
  year={2013}
}

@inproceedings{vectorCompression,
  title={Additive quantization for extreme vector compression},
  author={Babenko, Artem and Lempitsky, Victor},
  booktitle={CVPR},
  pages={931--938},
  year={2014}
}

@article{IMI,
  title={The inverted multi-index},
  author={Babenko, Artem and Lempitsky, Victor},
  journal={IEEE TPAMI},
  volume={37},
  number={6},
  pages={1247--1260},
  year={2014},
  publisher={IEEE}
}

@article{VQ,
  title={Accelerating large-scale inference with anisotropic vector quantization},
  author={Guo, Ruiqi and Sun, Philip and Lindgren, Erik and Geng, Quan and Simcha, David and Chern, Felix and Kumar, Sanjiv},
  journal={ICML},
  pages={3887--3896},
  year={2020},
  organization={PMLR}
}

@article{CacheLocality,
  title={Cache locality is not enough: High-performance nearest neighbor search with product quantization fast scan},
  author={Andr{\'e}, Fabien and Kermarrec, Anne-Marie and Le Scouarnec, Nicolas},
  journal={VLDB},
  volume={9},
  number={4},
  pages={12},
  year={2016}
}

@article{RaBitQ,
  title={RaBitQ: Quantizing High-Dimensional Vectors with a Theoretical Error Bound for Approximate Nearest Neighbor Search},
  author={Gao, Jianyang and Long, Cheng},
  journal={ACM SIGMOD},
  volume={2},
  number={3},
  pages={1--27},
  year={2024},
  //publisher={ACM New York, NY, USA}
}

@article{ScalableNN,
  title={Scalable nearest neighbor algorithms for high dimensional data},
  author={Muja, Marius and Lowe, David G},
  journal={IEEE TPAMI},
  volume={36},
  number={11},
  pages={2227--2240},
  year={2014},
  publisher={IEEE}
}

@misc{annoy,
  title   = {Annoy: Approximate Nearest Neighbors Oh Yeah},
  author  = {Bernhardsson, Erik},
  year    = {2013},
  howpublished = {\url{https://github.com/spotify/annoy}},
  note = {Accessed: 2025-01-21}
}

@article{K-means,
  title={Genetic K-means algorithm},
  author={Krishna, K and Murty, M Narasimha},
  journal={IEEE Transactions on Systems, Man, and Cybernetics, Part B (Cybernetics)},
  volume={29},
  number={3},
  pages={433--439},
  year={1999},
  publisher={IEEE}
}

@article{BlendHouse,
  title={BlendHouse: A Cloud-Native Vector Database System in ByteHouse},
  author={Niu, Zhaojie and Tian, Xinhui and Peng, Xindong and Chen, Xing},
  journal={IEEE ICDE},
  pages={4332--4345},
  year={2025},
  organization={IEEE}
}

@article{XN-Graph,
  title={Highly Efficient Disk-based Nearest Neighbor Search on Extended Neighborhood Graph},
  author={Zhang, Cheng and Wang, Jianzhi and Zhao, Wan-Lei and Xiao, Shihai},
  journal={ACM SIGIR},
  pages={2513--2523},
  year={2025}
}

@article{BAMG,
  title={BAMG: A Block-Aware Monotonic Graph Index for Disk-Based Approximate Nearest Neighbor Search},
  author={Li, Huiling and Xu, Jianliang},
  journal={arXiv preprint arXiv:2509.03226},
  year={2025}
}

@article{LSM-VEC,
  title={LSM-VEC: A Large-Scale Disk-Based System for Dynamic Vector Search},
  author={Zhong, Shurui and Mo, Dingheng and Luo, Siqiang},
  journal={arXiv preprint arXiv:2505.17152},
  year={2025}
}

@article{trim,
  title={TRIM: Accelerating High-Dimensional Vector Similarity Search with Enhanced Triangle-Inequality-Based Pruning},
  author={Song, Yitong and Zhang, Pengcheng and Gao, Chao and Yao, Bin and Wang, Kai and Wu, Zongyuan and Qu, Lin},
  journal={ACM SIGMOD},
  pages={1--26},
  year={2025}
}

@article{Gorgeous,
  title={Gorgeous: Revisiting the Data Layout for Disk-Resident High-Dimensional Vector Search},
  author={Yin, Peiqi and Yan, Xiao and Zhou, Qihui and Li, Hui and Li, Xiaolu and Zhang, Lin and Wang, Meiling and Yao, Xin and Cheng, James},
  journal={arXiv preprint arXiv:2508.15290},
  year={2025}
}

@article{icde_learned_functions,
  title={I/O efficient approximate nearest neighbour search based on learned functions},
  author={Li, Mingjie and Zhang, Ying and Sun, Yifang and Wang, Wei and Tsang, Ivor W and Lin, Xuemin},
  journal={IEEE ICDE},
  pages={289--300},
  year={2020}
}

@article{ei-lsh,
  title={EI-LSH: An early-termination driven I/O efficient incremental c-approximate nearest neighbor search},
  author={Liu, Wanqi and Wang, Hanchen and Zhang, Ying and Wang, Wei and Qin, Lu and Lin, Xuemin},
  journal={VLDBJ},
  volume={30},
  pages={215--235},
  year={2021}
}

@article{pqbf,
  title={PQBF: i/o-efficient approximate nearest neighbor search by product quantization},
  author={Liu, Yingfan and Cheng, Hong and Cui, Jiangtao},
  journal={CIKM},
  pages={667--676},
  year={2017}
}

@article{I-LSH,
  title={I-LSH: I/O efficient c-approximate nearest neighbor search in high-dimensional space},
  author={Liu, Wanqi and Wang, Hanchen and Zhang, Ying and Wang, Wei and Qin, Lu},
  journal={IEEE ICDE},
  pages={1670--1673},
  year={2019}
}

@article{spfresh,
  title={Spfresh: Incremental in-place update for billion-scale vector search},
  author={Xu, Yuming and Liang, Hengyu and Li, Jin and Xu, Shuotao and Chen, Qi and Zhang, Qianxi and Li, Cheng and Yang, Ziyue and Yang, Fan and Yang, Yuqing and others},
  journal={SOSP},
  pages={545--561},
  year={2023}
}

@article{zoom,
  title={Zoom: SSD-based vector search for optimizing accuracy, latency and memory},
  author={Zhang, Minjia and He, Yuxiong},
  journal={arXiv preprint arXiv:1809.04067},
  year={2018}
}

@article{DiskANNplus,
  title={DiskANN++: Efficient page-based search over isomorphic mapped graph index using query-sensitivity entry vertex},
  author={Ni, Jiongkang and Xu, Xiaoliang and Wang, Yuxiang and Li, Can and Yao, Jiajie and Xiao, Shihai and Zhang, Xuecang},
  journal={arXiv preprint arXiv:2310.00402},
  year={2023}
}

@article{LEANN,
  title={LEANN: A Low-Storage Vector Index},
  author={Wang, Yichuan and Liu, Shu and Li, Zhifei and Wu, Yongji and Mao, Ziming and Zhao, Yilong and Yan, Xiao and Xu, Zhiying and Zhou, Yang and Stoica, Ion and others},
  journal={arXiv preprint arXiv:2506.08276},
  year={2025}
}

@article{PipeANN,
  title={Achieving Low-Latency Graph-Based Vector Search via Aligning Best-First Search Algorithm with SSD},
  author={Guo, Hao and Lu, Youyou},
  journal={19th USENIX Symposium on Operating Systems Design and Implementation (OSDI 25)},
  pages={171--186},
  year={2025}
}

@article{PageANN,
  title={Scalable Disk-Based Approximate Nearest Neighbor Search with Page-Aligned Graph},
  author={Kang, Dingyi and Jiang, Dongming and Yang, Hanshen and Liu, Hang and Li, Bingzhe},
  journal={arXiv preprint arXiv:2509.25487},
  year={2025}
}

@inproceedings{MicroNN,
  title={{MicroNN}: An On-device Disk-resident Updatable Vector Database},
  author={Pound, Jeffrey and Chabert, Floris and Bhushan, Arjun and Goswami, Ankur and Pacaci, Anil and Chowdhury, Shihabur Rahman},
  booktitle={Companion of the 2025 International Conference on Management of Data},
  pages={608--621},
  year={2025}
}

@article{hy_ivf_flat,
  title={Billion-Scale Similarity Search Using a Hybrid Indexing Approach with Advanced Filtering},
  author={Emanuilov, Simeon and Dimov, Aleksandar},
  journal={Cybernetics and Information Technologies},
  volume={24},
  number={4},
  pages={45--58},
  year={2024}
}

@article{CrackIVF,
  title={Cracking Vector Search Indexes},
  author={Mageirakos, Vasilis and Wu, Bowen and Alonso, Gustavo},
  journal={arXiv preprint arXiv:2503.01823},
  year={2025}
}

@article{margo,
  title={Select Edges Wisely: Monotonic Path Aware Graph Layout Optimization for Disk-Based ANN Search},
  author={Yue, Ziyang and Zheng, Bolong and Xu, Ling and Xu, Kanru and Zhang, Shuhao and Du, Yajuan and Gao, Yunjun and Zhou, Xiaofang and Jensen, Christian S},
  journal={PVLDB},
  volume={18},
  number={11},
  pages={4337--4349},
  year={2025},
}

@article{DGAI,
  title={DGAI: Decoupled On-Disk Graph-Based ANN Index for Efficient Updates and Queries},
  author={Lou, Jiahao and Yu, Quan and Gong, Shufeng and Yu, Song and Zhang, Yanfeng and Yu, Ge},
  journal={arXiv preprint arXiv:2510.25401},
  year={2025}
}

@article{graphSurvey,
  title={A comprehensive survey and experimental comparison of graph-based approximate nearest neighbor search},
  author={Wang, Mengzhao and Xu, Xiaoliang and Yue, Qiang and Wang, Yuxiang},
  journal={arXiv preprint arXiv:2101.12631},
  year={2021}
}

@article{gao2025trae,
  title={Trae agent: An llm-based agent for software engineering with test-time scaling},
  author={Gao, Pengfei and Tian, Zhao and Meng, Xiangxin and Wang, Xinchen and Hu, Ruida and Xiao, Yuanan and Liu, Yizhou and Zhang, Zhao and Chen, Junjie and Gao, Cuiyun and others},
  journal={arXiv preprint arXiv:2507.23370},
  year={2025}
}

@online{ZillizCloud,
  author       = {Cal Huang},
  title        = {Migrating from S3 Vectors to Zilliz Cloud: Unlocking the Power of Tiered Storage},
  year         = {2025},
  month        = {Oct},
  day          = {23},
  url          = {https://zilliz.com/blog/migrating-from-s3-vectors-to-zilliz-cloud-unlocking-the-power-of-tiered-storage},
  note         = {Blog post, Zilliz},
}

@online{AmazonS3,
  author       = {Channy Yun},
  title        = {Introducing Amazon S3 Vectors: First cloud storage with native vector support at scale (preview)},
  year         = {2025},
  month        = {Jul},
  day          = {15},
  url          = {https://aws.amazon.com/cn/blogs/aws/introducing-amazon-s3-vectors-first-cloud-storage-with-native-vector-support-at-scale/},
  note         = {Blog post, AWS News Blog},
}

@online{turbopuffer,
  author       = {Turbopuffer Inc.},
  title        = {Turbopuffer — Serverless Vector \& Full-Text Search Built from First Principles on Object Storage},
  year         = {2025},
  url          = {https://turbopuffer.com/},
  note         = {Company website / product page},
}

@inproceedings{tutorial1,
  title={Vector database management techniques and systems},
  author={Pan, James Jie and Wang, Jianguo and Li, Guoliang},
  booktitle={Companion of the 2024 International Conference on Management of Data},
  pages={597--604},
  year={2024}
}

@article{tutorial2,
  title={New trends in high-d vector similarity search: al-driven, progressive, and distributed},
  author={Echihabi, Karima and Zoumpatianos, Kostas and Palpanas, Themis},
  journal={PVLDB},
  volume={14},
  number={12},
  pages={3198--3201},
  year={2021},
  publisher={VLDB Endowment}
}

@article{Quake,
  title={Quake: Adaptive Indexing for Vector Search},
  author={Mohoney, Jason and Sarda, Devesh and Tang, Mengze and Chowdhury, Shihabur Rahman and Pacaci, Anil and Ilyas, Ihab F and Rekatsinas, Theodoros and Venkataraman, Shivaram},
  journal={arXiv preprint arXiv:2506.03437},
  year={2025}
}

@inproceedings{BalancedCluster,
  author    = {Hongfu Liu and Ziming Huang and Qi Chen and Mingqin Li and Y. Fu and Lintao Zhang},
  year      = {2018},
  title     = {Fast Clustering with Flexible Balance Constraints},
  booktitle = {2018 IEEE International Conference on Big Data (Big Data)}
}

@inproceedings{BalancedKMeans,
  title={Balanced k-means for clustering},
  author={Malinen, Mikko I and Fr{\"a}nti, Pasi},
  booktitle={Joint IAPR international workshops on statistical techniques in pattern recognition (SPR) and structural and syntactic pattern recognition (SSPR)},
  pages={32--41},
  year={2014},
  organization={Springer}
}

@article{CloudNativeVS,
  title   = {Cloud-Native Vector Search: A Comprehensive Performance Analysis},
  author  = {Li, Zhaoheng and Ding, Wei and Huang, Silu and Wang, Zikang and Lin, Yuanjin and Wu, Ke and Park, Yongjoo and Chen, Jianjun},
  journal = {arXiv preprint arXiv:2511.14748},
  year    = {2025}
}

@article{UNIFY,
  title={UNIFY: Unified Index for Range Filtered Approximate Nearest Neighbors Search},
  author={Liang, Anqi and Zhang, Pengcheng and Yao, Bin and Chen, Zhongpu and Song, Yitong and Cheng, Guangxu},
  journal={PVLDB},
  year={2025}
}

\end{document}